\documentclass[vecphys]{svmult}

\usepackage{makeidx}         
\usepackage{graphicx}        
\usepackage{multicol}        
\usepackage[bottom]{footmisc}

\makeindex             

\begin{document}

\title*{`Disc-jet' coupling in black hole X-ray binaries and active
  galactic nuclei}

\author{Rob Fender\inst{1}}

\institute{School of Physics \& Astronomy, University of Southampton, SO17 1BJ, UK
\texttt{rpf@phys.soton.ac.uk}}

\maketitle

\abstract{ In this chapter I will review the status of our
phenomenological understanding of the relation between accretion and
outflows in accreting black hole systems. This understanding arises
primarily from observing the relation between X-ray and longer
wavelength (infrared, radio) emission. The view is necessarily a
biased one, beginning with observations of X-ray binary systems, and
attempting to see if they match with the general observational
properties of active galactic nuclei.  }
\label{intro}

Black holes are amongst the most esoteric objects conceived of by
man. They do not however lurk only in our imaginations or at the
fringes of reality but appear, as far as we can tell from the
objective interpretation of astrophysical observations, to play a
major role in the history of the entire universe.

\begin{figure}
\includegraphics[width=12cm]{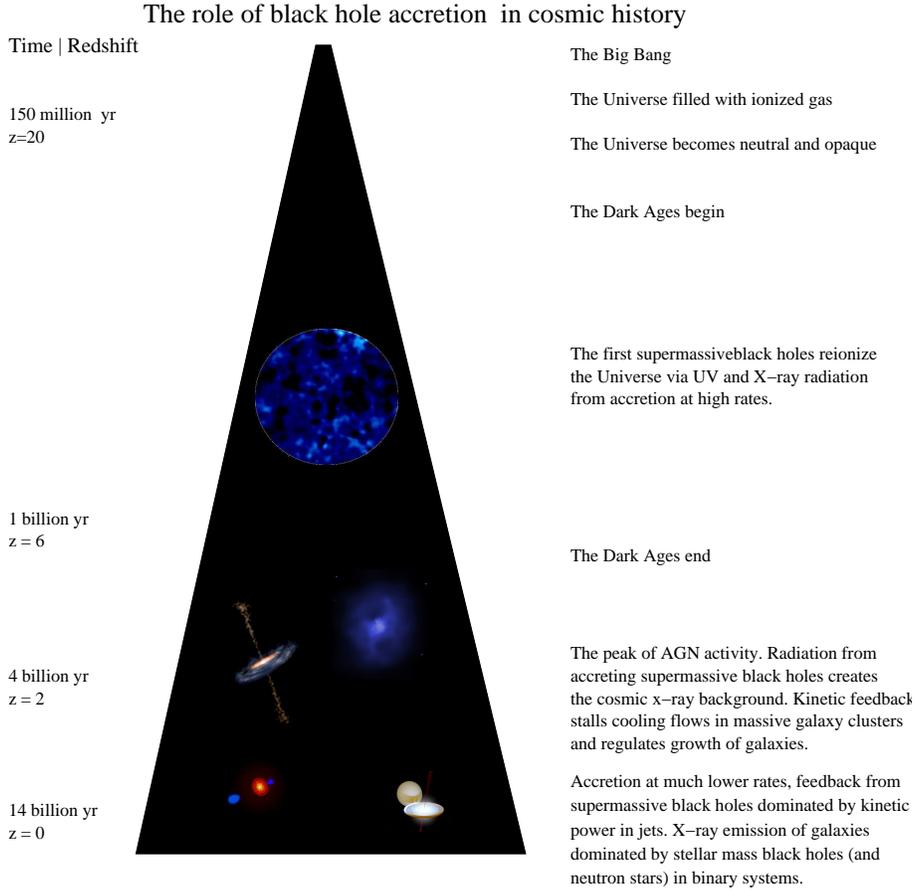}
\caption{The role of feedback from black hole accretion, in the form
  of both radiation and kinetic energy, has played a key role in the
  evolution of our Universe.}
\label{cosmos}
\end{figure}

In particular, feedback from accreting black holes, in the form of
both radiation and kinetic energy (ie. jets and winds) has been a key
element throughout cosmic evolution. This is summarised in
Fig. \ref{cosmos}, and some key phases can be identified as:

\begin{itemize}
\item{{\bf Reionization:} By redshift of $z \sim 10$, less than a
  billion years after the big bang, UV and X-ray radiation from the
  first accreting black holes (as well as the first stars) undoubtedly
  played a key role in ending the so-called `Dark Ages' and reionizing
  the universe.}
\item{{\bf The epoch of Active Galactic Nuclei:} Following
  reionization, supermassive black holes at the centres of Active
  Galactic Nuclei (AGN) grew rapidly, probably fed by galactic
  mergers. The peak of AGN activity occurred around a redshift of $z
  \sim 2$, around 4 billion years after the big bang. Radiation from
  the accretion flows around these AGN formed what we now observe as
  the cosmic X-ray background. Kinetic feedback from AGN also acted to
  stall and reheat cooling flows at the centres of galaxies, and to
  regulate the growth of galaxies.}
\item{{\bf The epoch of stellar mass black holes:} In the nearby
  Universe feeding of the supermassive black holes has declined, and
  their feedback to the ambient medium is dominated by kinetic
  power. The X-ray luminosity of galaxies is now dominated by
  accretion onto black holes of mass $M_{\rm BH} \sim 10_{\odot}$ in
  binary systems (as well as neutron stars).}
\end{itemize}

Black holes, it seems, despite their oddness, are a key formative
component of the Universe. It is essential for our understanding not
only of the history of the Universe to this point, but also of the
future evolution of the universe, that we understand how these objects
behave. In fact, they do one thing, and one thing only, of major
significance:{\footnote{Not counting information trapping on their
    surfaces (e.g. \cite{hooft})}} they convert gravitational
potential energy to radiation and kinetic energy, which feeds back
into the universe.

This accretion process is in principle quite simple, as outlined in
the next section. However, observationally we find that it has its
subtleties and nuances, which manifest themselves most clearly in how
the black hole distributes its feedback between radiation and kinetic
power. The rest of this chapter explores what we can learn about these
idiosyncrasies of black hole accretion by studying low-mass ($M < 20
M_{\odot}$), rapidly varying, black holes in binary systems in our
galaxy, and how we might apply that to understanding how supermassive
black holes have helped shape the observable Universe.

In addition to the flow of energy and feedback, which are the foci of
this review, the physics associated with jet formation and the
associated particle acceleration, as well potential tests of general
relativity associated with studying black hole accretion are all
extremely interesting astrophysical topics in their own right. See
suggestions for further reading at the end of the chapter.

\section{Simple physical theory}

The maximum energy release associated with the accretion of matter
from infinity to a body of mass $M$ and radius $R$ is given by $G M /
R$. It is the ratio $M / R$ which determines the efficiency of the
accretion process.

\begin{figure}
\includegraphics[width=12cm]{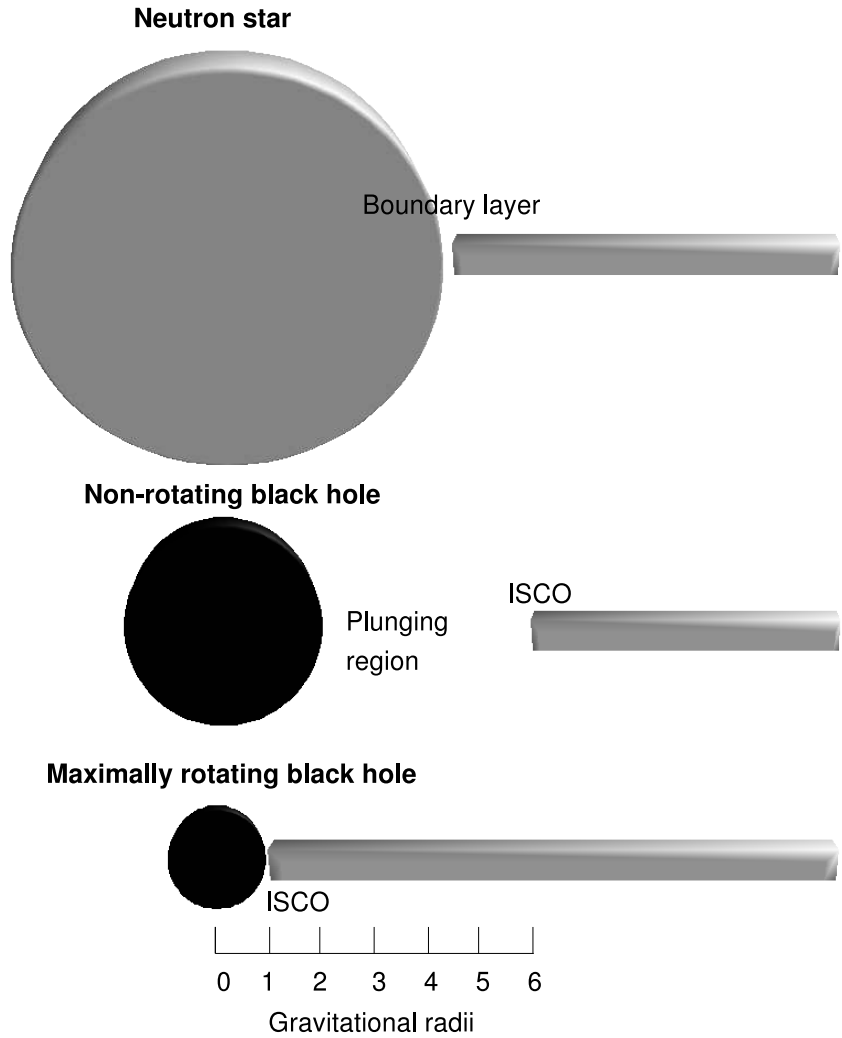}
\caption{Relative sizes of a neutron star, non-rotating and maximally
  rotating black hole in units of gravitational radii $R_{\rm G} = GM
  / c^2$. The figure is correct for black holes of any mass.  A
  neutron star of 1.4 M$_{\odot}$ mass and 10 km radius has a surface
  at about 4.8 $R_{\rm G}$. The accretion disc may extend all the way
  to the surface where there will be a boundary layer between the
  inner disc edge and the surface, which will be rotating more
  slowly. For a non-rotating black hole the event horizon will lie at
  $2R_{\rm G}$ and there will be an innermost stable circular orbit
  (ISCO) at $6R_{\rm G}$. Inside of the ISCO matter will plunge across
  the event horizon. For a maximally rotating black hole the event
  horizon and ISCO lie at $1R_{\rm G}$. Most black holes will be
  somewhere between these two extremes. If we naively assume radiation
  only from a disc-like radiatively efficient accretion flow, we can
  see that accretion onto a neutron star can be more efficient than
  that onto a non-rotating black hole, but that accretion onto a black
  hole with significant spin can be the most efficient process.}
\label{sizes}
\end{figure}

A non-rotating, or {\emph Schwarzschild} black hole has an event horizon 
\index{event horizon} at 

\[
r_s = 2 G M / c^2 = 2 r_g
\]

where $r_g$ is referred to as the gravitational radius.

This linear dependence on black hole mass means that all non-rotating
black holes have the same maximum potential accretion efficiency,
regardless of their mass. There is an {\emph innermost stable circular
  orbit}, or ISCO\index{ISCO}, around such black holes, within which matter will
plummet rapidly across the event horizon, and this is at $3 r_S$, also
independent of mass.

Spinning black holes have smaller event horizons and ISCOs than
non-rotating black holes: a maximally rotating or {\emph Maximal Kerr}
black hole has an event horizon and ISCO both at $r_g = \frac12
r_s$. This means that for a black hole of any mass, the ratio $M / R$
is the same to within a factor of six. This ratio is dwarfed by the
enormous range of masses of black holes we have observed in the
universe, from $\leq 10 M_{\odot}$ to $\geq 10^9 M_{\odot}$. Beyond
mass and spin, black holes possess only one more property, that of
electric charge. Given the lack of observation of large-scale charged
objects in the universe, it is assumed that black holes are
electrically neutral. Fig. \ref{sizes} summarizes the relative sizes of
neutron stars, non-rotating and maximally rotating black holes.

This amazing scale\index{scaling} invariance of the theoretical accretion efficiency
coupled with the inability of a black hole to possess any other
distinguishing characteristics drives us to speculate that the process
of accretion onto black holes of all scales should be very similar.

\begin{figure}
\includegraphics[width=8.5cm, angle=-90]{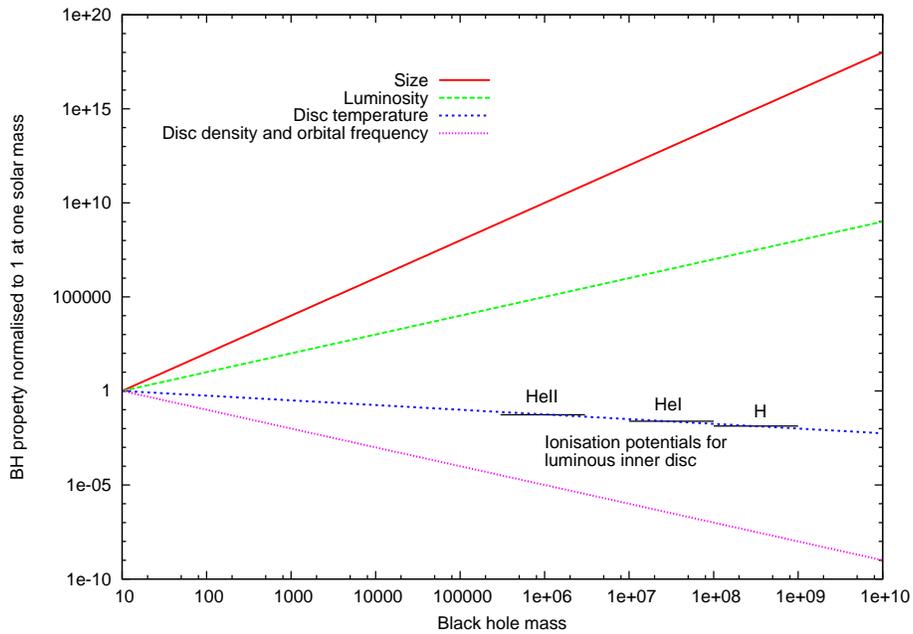}
\caption{The simple scaling\index{scaling} of size (both of event horizon and inner
  disc), luminosity, disc temperature, disc orbital frequency and disc
  density (also mean density within event horizon) with black hole
  mass, for black holes accreting radiatively efficiently at the same
  Eddington ratio (the combination of which requirements probably
  means an Eddington ratio $\geq 0.01$) and with the same spin. Also
  indicated are the ionisation potentials for H and both HeI and HeII,
  based on a temperature of 1 keV for an Eddington rate $10_{\odot}$ 
  black hole (i.e. the temperature scale is the actual temperature in keV).}
\label{scalings}
\end{figure}

However, the physical conditions in the innermost regions of the
accretion flow are {\emph not} expected to be identical. The fixed $M /
R$ ratio necessarily implies that the density of the inner accretion
flow (and in fact of the black hole itself) must decrease with
increasing mass.  Put very simplistically, unless there are strong
changes in accretion geometry with mass, then the surface area through
which matter with significant angular momentum will accrete will scale
as $M^2$. So for two black holes, one `stellar mass' ($\sim 10
M_{\odot}$) and one `supermassive' ($\sim 10^9 M_{\odot}$), accreting
at the same Eddington ratio, which scales linearly with mass, the
density of the accreting matter should scale as $\rho \propto
\dot{m}/M^2 \propto M^{-1}$, i.e. a factor of $10^7$ more dense in the
case of the stellar mass black hole. Similarly, the effective
temperature of the accretion disc varies with mass. This can be
illustrated simply as follows: for the same Eddington ratio the
luminosity released in radiatively efficient accretion $L \propto M$,
and is emitted over a disc area which scales as $M^2$. For black body
radiation $L \propto A T^4$ (where $A$ is emitting area and $T$
temperature), and so we find $T \propto M^{-1/4}$. Therefore,
accretion discs in the most luminous AGN should be 100 times cooler
than those in the most luminous X-ray binaries \cite{ss76}. In Fig. \ref{scalings} we present the expected simple scaling of
size (both of the black hole and inner accretion disc), luminosity,
disc temperature, disc density (also mean density of the matter within
the event horizon) and also orbital frequency as a function of mass,
for black holes accreting radiatively efficiently at the same
Eddington ratio, and with the same spin. As we shall see later in this
article, the requirement for radiative efficiency probably implies an
Eddington ratio $\geq 0.01$.

How the energy density and organisation of the magnetic field in the
accretion flow, which are likely to be important both for effective
viscosity and jet formation, vary with black hole mass, is less well
understood.

\section{Observations of black hole X-ray binaries}

More than 20 objects consistent with black holes of mass $\sim 10
M_{\odot}$ have been identified in X-ray binary systems within our
galaxy (e.g. \cite{mcrem06}). This population may be the
tip of an iceberg of $\sim 10^8$ stellar-mass black holes within our
galaxy (i.e. a far larger total black hole mass than that associated
with the $\sim 10^6 M_{\odot}$ black hole, Sgr~A*\index{Sgr~A*}, at our galactic
centre).  These binary systems show semi-regular outbursts in which
they temporarily brighten across the electromagnetic spectrum by many
orders of magnitude.  The origin of these outbursts is believed to be
a disc instability driven by the ionization of hydrogen above a given
temperature (e.g. \cite{fkr}). However while describing
general outburst trends, there are many difficulties in explaining all
the observational characteristics. When comparing to AGN, we imagine
that X-ray binaries near the peaks of their outbursts correspond to
the high Eddington-ratio systems such as Quasars, and that when in
quiescence (the most common phase for most systems) they correspond to
low luminosity AGN (LLAGN) such as Sgr~A* at the centre of our own
galaxy.

The evolution of two outbursts from the binary GX~339-4\index{GX~339-4} are
illustrated in Fig. \ref{dunn}. In very brief summary, most black hole
X-ray binaries spend most of their time in a `quiescent' state with
X-ray luminosities as low as $\sim 10^{30}$ erg s$^{-1}$ ($\leq
10^{-9} L_{\rm Edd}$; e.g. \cite{garcia}). Mass transfer from
the companion star, usually via Roche Lobe overflow, progresses at a
higher rate than that at which matter is centrally accreted onto the
black hole (or neutron star), and so the mass (and temperature) of the
disc increase. At some point the effective viscosity of the disc
increases, perhaps due to the hydrogen ionization instability mentioned
above, and the matter in the disc rapidly drains towards the central
accretor. During this phase the central accretion rate is much higher
than the time-averaged mass accretion rate from the companion star,
and the source becomes very luminous, often approaching the Eddington
luminosity 
\footnote{The {\emph Eddington Luminosity} corresponds to the luminosity
  at which the outwards radiation force from accretion balances the
  inward force of gravity. For spherical accretion of hydrogen it is
  approximately $L_{\rm Edd} \sim 1.4 \times 10^{38} (M/M_{\odot})$
  erg s$^{-1}$}.  

An early stage in the outburst the source X-ray spectrum begins to
soften (motion to the left in the HIDs\index{Hardness-Intensity Diagram (HID)} in Fig. 3).  Within a few weeks
or months, when some significant fraction of the disc mass has been
accreted, the disc cools, and the source returns towards quiescence
{\footnote{It is worth bearing in mind that two of the sources we use
    most in our studies of black hole binaries do not really fit this
    pattern: GX~339-4 doesn't ever really settle into extended
    quiescent periods (see Fig. \ref{dunn}), and Cyg~X-1 is not
    accreting via simple Roche lobe overflow.}  In doing so it makes
  the X-ray spectral transition away from the soft state at a lower
  luminosity than that at which it entered the state. This results in
  a hysteretical track in the HID.  For more details on black hole
  outbursts see Chaps. 3,4 and 6 in this volume.

\begin{figure}
\includegraphics[width=12cm]{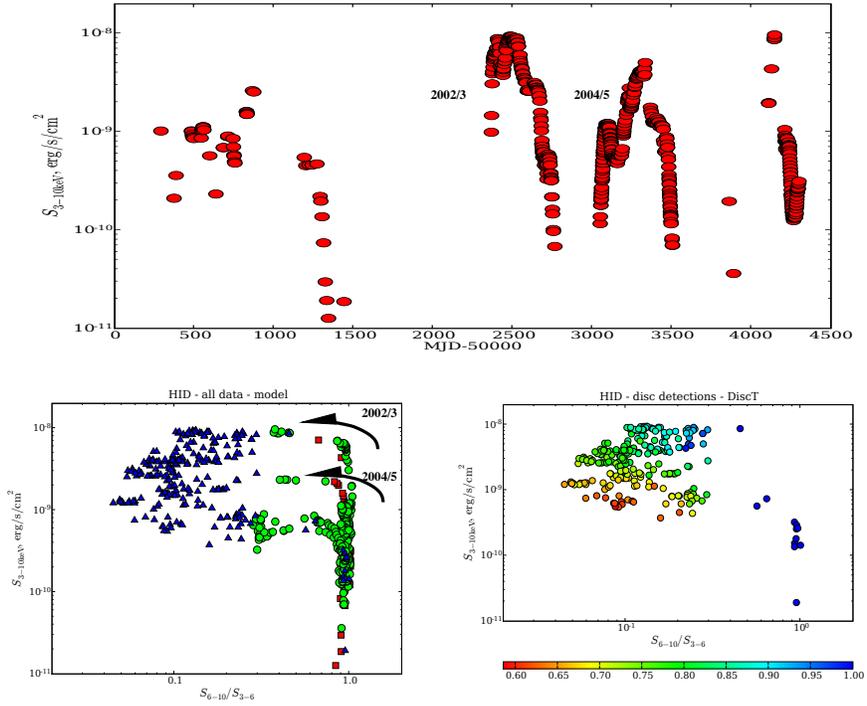}
\caption{Outbursts of the black hole binary GX~339-4. {\bf Top panel:}
  the X-ray light curve of GX~339-4 as measured with the PCA
  instrument onboard the RXTE satellite. Gaps in the data coverage
  between major outbursts usually indicate low / quiescent flux
  levels. The outbursts in 2002/3 and 2004/5 are covered in the
  greatest detail and a represented in Hardness-Intensity Diagrams
  (HIDs) in the lower panels.  In these HIDs the abscissae (x-axes)
  indicate x-ray `colour' or hardness ratio, whereas the ordinates
  (y-axes) indicate luminosity. In the {\bf left panel} the time
  evolution of the two separate outbursts are indicated; triangles
  indicate those spectra where a strong accretion disc (black
  body-like) component was required in the spectral fit. Two different
  outbursts are overplotted -- both show hysteretical\index{hysteresis} patterns of
  behaviour, in that the transition from hard $\rightarrow$ soft
  states occurs at higher luminosities than the soft $\rightarrow$
  hard transition, but they also clearly differ in that the earlier
  outburst reached higher a higher luminosity in the initial
  stages. The {\bf right panel} indicates the variation of fitted
  accretion disc temperature (colour scale) in the soft X-ray state,
  decreasing with luminosity. Adapted from\cite{dunn}.}
\label{dunn}
\end{figure}

\subsection{Relations between accretion `state' and radio emission}

Most of our insight into the relation between modes/rates of accretion
and the type and power of any associated feedback (also known as the
\index{disc--jet coupling}
`disc--jet' coupling {\footnote{This expression now has confusing
connotations because people sometimes take the word `disc' in this
context to mean only the geometrically thin, optically thick component
of the accretion flow}}) come from (near-)simultaneous radio (also
sometimes infrared) and X-ray observations of rapidly varying
systems. The radio (and, often, infrared) emission is assumed to arise via synchrotron
emission in a jet-like outflow, and the X-ray emission to be a tracer
of the accretion flow (rate, geometry, temperature, even
composition). It has also been suggested that a significant fraction
of the X-ray emission in some states may arise in the jet (see Chap.
6).

\begin{figure}
\includegraphics[width=9cm, angle=-90]{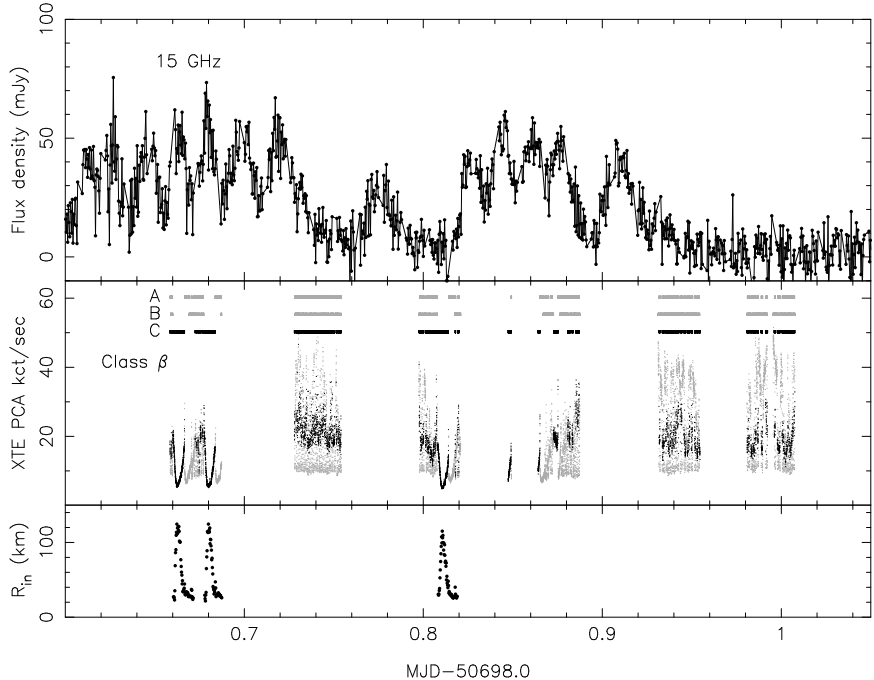}
\caption{Radio--X-ray coupling in the binary black hole system GRS~1915+105.  
  The {\bf top panel} shows radio monitoring at 15 GHz from the
  Ryle Telescope, the {\bf middle panel} shows the X-ray flux as measured by
  RXTE, with the light curve subdivided into `soft' (A and B) and
  `hard' (C) X-ray states. The radio oscillation events, which we
  think are associated with individual relativistic ejection events,
  only occur when long hard (state C) dips are followed by rapid
  transitions to soft states (e.g. the first and third phases of X-ray
  coverage) and not when more rapid fluctuations occur, albeit at the
  same luminosity. While GRS~1915+105 is a complex case, the general
  pattern of behaviour -- major ejection events during state
  transitions -- is consistent with other accreting black hole
  binaries. From \cite{klein2002}.}
\label{k-w}
\end{figure}

A good example of coupling between X-ray state and radio emission is
given in Fig. \ref{k-w}. In this figure several hours of overlapping
radio and X-ray observations of the powerful jet source and black hole
binary GRS~1915+105\index{GRS~1915+105} are presented (from \cite{klein2002}). The
radio emission shows clear phases of oscillatory behaviour during
periods of strong dips in the hard X-ray light curve, and far less
activity when the X-rays, although at the same luminosity, are not
showing the long hard state dips. Similar, maybe identical, patterns
are observed at widely separated epochs, demonstrating a clear and
repeating pattern of accretion:outflow coupling.

\subsection{Towards unified models for accretion:ejection coupling}

Based on more than a decade of X-ray and radio observations, attempts
were made in the past few years to find simple and unified patterns
for the accretion -- outflow (`disc--jet') coupling in black hole
X-ray binaries. Our analysis is heavily based upon the assumption that
radio emission is associated with jet-like outflows, something argued
in more detail in e.g. \cite{fender2006} and by many other authors. In terms
of relations to X-ray states, \cite{fender1999} demonstrated that
during the high/soft X-ray state of the binary GX~339-4 the radio
emission was suppressed with respect to the hard state at comparable
luminosities. In fact this phenomena had been observed more than two
decades earlier by \cite{tanan} in the case of Cygnus~X-1.
Around the same time, \cite{diana,corbel2000,corbel2003} 
demonstrated that while in the hard X-ray state
(right hand side of the the HID) the same binary, GX~339-4\index{GX~339-4}, repeatedly
displayed a correlation of the form

\[
L_{\rm GHz radio} \propto L_{\rm soft X-ray}^b
\]

where $b = 0.7 \pm 0.1$. \cite{fender2001}
demonstrated that steady, flat-spectrum radio emission
(spectral index $\alpha = \Delta \log S_{\nu} / \Delta \log \nu \sim
0$) such as observed from GX~339-4 in the hard X-ray state, was
observed from {\emph all} hard state black hole binaries (this
characteristic post-outburst radio behaviour had been noted by
e.g. \cite{han} but not clearly associated with X-ray
state).

In a wider-ranging study, \cite{gfp} found that another
binary V404~Cyg (GS~2023+338)\index{V404~Cyg} displayed the same correlation ($b \sim
0.7$) {\emph and} the same normalisation, within uncertainties. Radio
and X-ray measurements for several other black hole binaries in the
hard state were also consistent with the same `universal' relation,
and repeated suppression of the radio emission in softer X-ray states
of Cygnus~X-1 was reconfirmed. More recently however, several hard
state black hole binaries have been discovered which are underluminous
in the radio band (see \cite{gallo2007}, and Chap. 4 in this volume
for more discussion).

These hard-state, flat-spectrum jets are in fact very powerful, and
not just an interesting sideshow to the main event of X-ray production
\cite{fender2001}. In fact the combination of two non-linear couplings,
$L_{\rm radio} \propto L_{\rm X}^{0.7}$ (observed)
\index{empirical correlations} and $L_{\rm radio}
\propto P_{\rm jet}^{1.4}$ (theoretical, where $P_{\rm jet}$ is total
jet power, but also observed in \cite{koerdingfendermigliari},
implies that $P_{\rm jet} \propto L_{\rm X}^{0.5}$, and that as the
X-ray luminosity of a source declines the jet may come to dominate
over radiation in terms of feedback from the accretion process
\cite{fgj}. The key goal then was to determine the
normalization for the jet power, which a variety of methods has
established as being comparable to the X-rays luminosity at high
Eddington ratios, dominating at lower Eddington ratios (e.g. \cite{gallo2005,heinzgrimm}).
The current consensus is therefore
that {\emph in the hard X-ray state black hole binaries produce powerful,
flat-spectrum relatively steady jets, whose strength correlates in a
(near-)universal way with the X-ray luminosity}.

However, really spectacular radio ejection events, during which
relativistic, sometimes apparently superluminal radio components were
observed to propagate away from the binary, were also known
(e.g. \cite{mirabel,hjellmingrupen}).  These jets
also carry away a large amount of power (see \cite{fbg}
and references therein).  How did these relate to the steady
radio emission in the hard state, and the apparently suppressed radio
emission in the soft state ?

Careful examination of X-ray data compared to radio monitoring and
imaging observations revealed the answer: such bright ejection events
occurred during the transition from hard to soft X-ray
states. Combining our knowledge of the hysteretical outburst behaviour
of black hole binaries (see Chap. 3) with these insights
into the coupling to radio emission, allowed a first attempt at a
`unified model' for the disc--jet coupling, which was put forward by
\cite{fbg}, see Fig. \ref{fbg}.

\begin{figure}
\includegraphics[width=12cm]{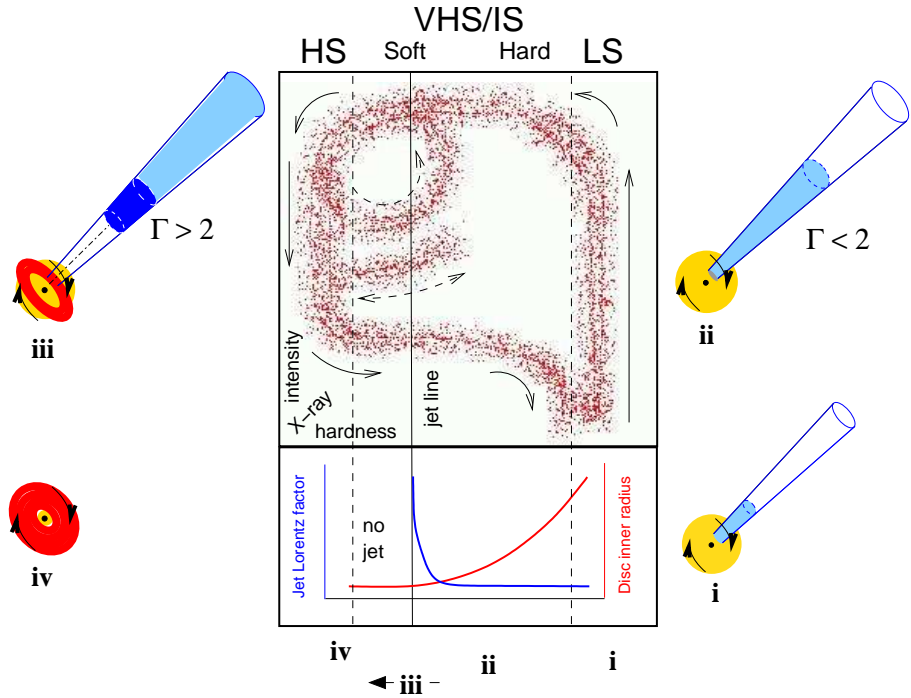}
\caption{A schematic of the simplified model for the jet-disc coupling
  in black hole binaries presented by \cite{fbg}. 
  The {\bf central box panel} represents an X-ray hardness-intensity
  diagram (HID); 'HS' indicates the `high/soft state', 'VHS/IS'
  indicates the 'very high/intermediate state' and 'LS' the 'low/hard
  state'. In this diagram, X-ray hardness increases to the right and
  intensity upwards. The {\bf lower panel} indicates the variation of the
  bulk Lorentz factor of the outflow with hardness -- in the LS and
  hard-VHS/IS the jet is steady with an almost constant bulk Lorentz
  factor $\Gamma < 2$, progressing from state {\bf i} to state {\bf
    ii} as the luminosity increases. At some point -- usually
  corresponding to the peak of the VHS/IS -- $\Gamma$ increases
  rapidly producing an internal shock in the outflow ({\bf iii})
  followed in general by cessation of jet production in a
  disc-dominated HS ({\bf iv}). At this stage fading optically thin
  radio emission is only associated with a jet/shock which is now
  physically decoupled from the central engine.  As a result the solid
  arrows indicate the track of a simple X-ray transient outburst with
  a single optically thin jet production episode. The dashed loop and
  dotted track indicate the paths that GRS~1915+105 and some other
  transients take in repeatedly hardening and then crossing zone {\bf
    iii} -- the 'jet line' -- \index{jet line}from left to right, producing further
  optically thin radio outbursts. Sketches around the outside
  illustrate our concept of the relative contributions of jet (blue),
  'corona' (yellow) and accretion disc (red) at these different
  stages.}
\label{fbg}
\end{figure}

Since this model was first proposed, nearly five years ago, we have
repeatedly attempted to test whether or not any of the basic empirical
relations was wrong. In particular, we have sought to confirm that the
major radio outbursts always occur during the hard $\rightarrow$ soft
transitions (i.e. left to right motion at the top of the HID\index{Hardness-Intensity Diagram (HID)}). In a
study of over 16 black hole outbursts (compared to four in \cite{fbg}) 
we (\cite{fhb09}) have
found no clear exceptions to this. In fact in revisiting the case of
the black hole transient XTE~J1859+226\index{XTE~J1859+226} we find strong evidence that
all of five recorded radio flare events (which we associate with an
ejection) occurred at $\sim$ same X-ray hardness (Fig. \ref{1859}).

\begin{figure}
\includegraphics[width=12cm, angle=0]{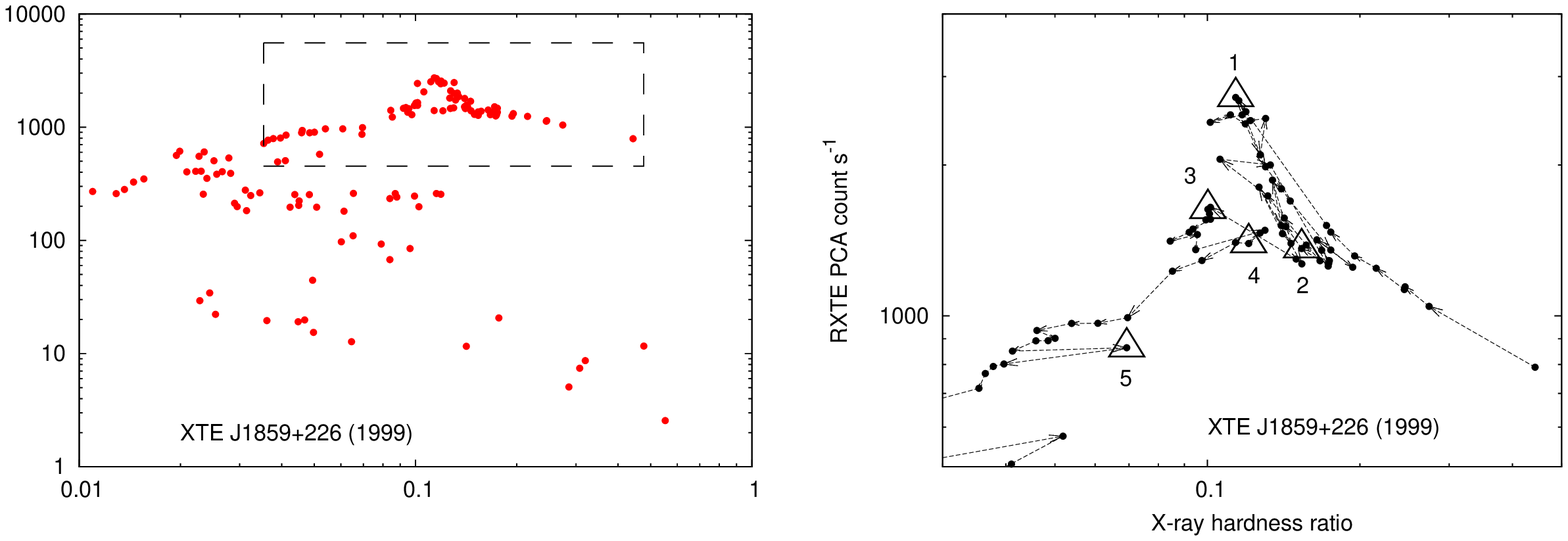}
\caption{Moments of five radio flare ejection events from the black
  hole binary XTE~J1859+226 (from \cite{brocksopp}) as a function
  of position in the HID. The {\bf left} panel indicates the overall
  HID for the source which travelled in a generally anti-clockwise
  direction in the figure, typical for such outbursts. In the {\bf
    right} panel we focus on the region (indicated by a box in the
  left panel) around the time of the radio ejections. Triangles
  indicate the estimated moments of ejection, and arrows indicate the
  temporal evolution in the HID. It is clear that all five radio
  ejection events took place at approximately the same hardness. Most
  notable is the last event, number 5, which is associated with a
  brief excursion back to harder states even as the source is in a
  softening trend. These date support the idea of a near-constant jet
  line on the upper branch of the HID, at least within one outburst
  loop. From \cite{fhb09}.}
\label{1859}
\end{figure}

While \cite{fbg} attempted to provide a simple
physical interpretation of the empirical relations, several more
detailed theoretical models have arisen based upon the emerging
phenomenology (e.g. \cite{ferreira,machida}; see also Chap. 9). A very interesting
combined observational / theoretical result was also presented by
\cite{mmf} who analysed optical and X-ray
observations of the hard state black hole XTE~J1118+480\index{XTE~J1118+480}. What they
found (see also \cite{kanbach,hynes}) was that the
very rapid correlated optical and X-ray variability in this source
could be explained by a strong synchrotron component in the optical,
which arose in a jet / coronal outflow which dominated the feedback or
accretion energy in this state -- the same conclusion as reached from
radio studies (see above).

\cite{fbg} only really connected the X-ray
spectral properties and long-term (i.e. hours and longer) evolution to
the state of the jet. Since then several groups have been trying to
see if the moment of ejection may be better determined by examining
the timing properties, as measured by X-ray power spectra (see also
Chap. 3). Certainly the hard to soft transition at the top
of the HID is both the time when the major ejections occur, and the
region when dramatic changes in the power spectra, including the
strongest QPOs, are observed (see \cite{klein2008}).
In Fig. \ref{zones} we plot X-ray r.m.s. variability as a function of
X-ray colour for three sources in outburst.  We indicate the moments
of bright radio flares (probably associated with relativistic ejection
events). Dips in the r.m.s.--colour relation, roughly indicated in the
figures, have been associated with the sharpest changes in the X-ray
timing properties, so are they related directly to the ejections ? The
result of the comparison is tantalising: in the case of XTE~J1550-564\index{XTE~J1550-564}
the bright radio flare occurred between the low-r.m.s. `zone' and an
earlier very sharp dip, and in the case of XTE~J1859+226 the five
radio flares all occurred around the `zone'. However, in GX~339-4, the
radio flare reported by \cite{gallo2004} occurred days {\emph before}
the `zone'.  Currently therefore the relation of major ejections to
source timing properties remains unclear.

\begin{figure}
\includegraphics[width=10cm, height=16cm, angle=0]{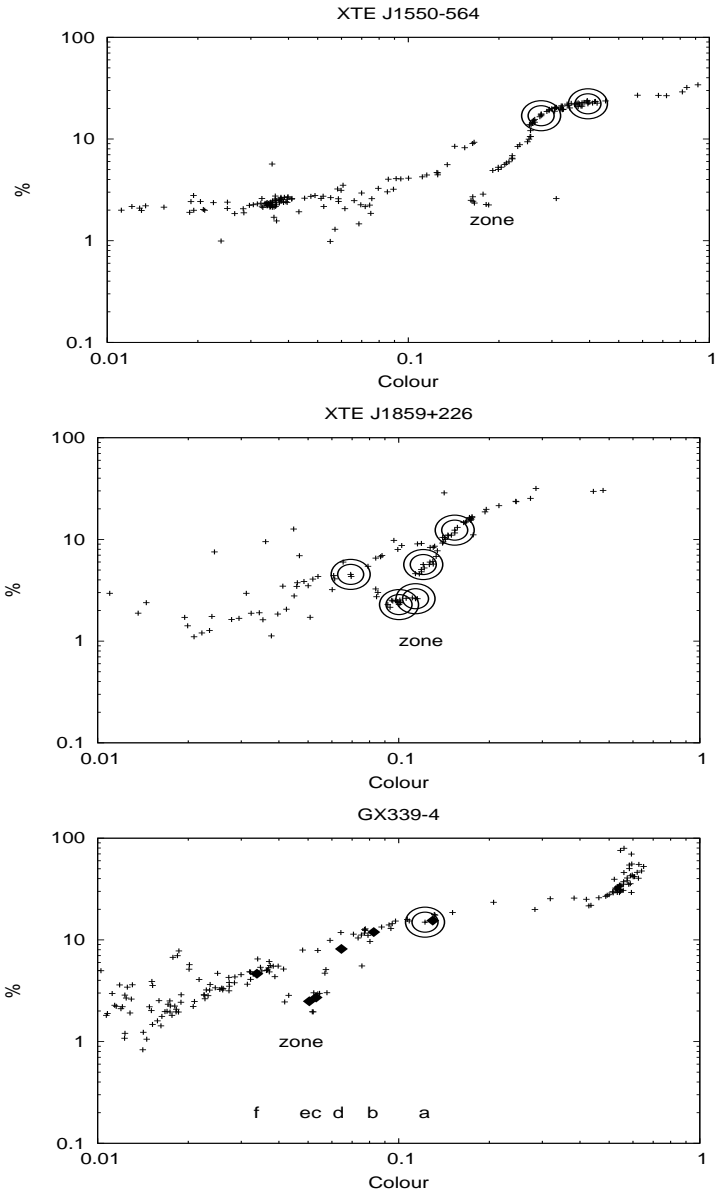}
\caption{Location of observed radio flare events (probably connected
  with relativistic ejections) in the hardness-r.m.s. diagram (which
  does not show the same hysteresis as the HID). Dramatic drops in
  X-ray r.m.s. are associated in many cases with the sharpest changes
  in X-ray power spectra during the overall hard $\rightarrow$ soft
  state transition, and have been speculatively linked with the
  relativistic ejections. In the case of XTE~J1550-564 ({\bf top panel}) the
  ejection appears to occur at the beginning of the drop to the
  low-r.m.s. `zone', \index{low-r.m.s. zone} and in XTE~J1859+226 ({\bf middle panel}) the sequence
  of five ejection events all happen close to the `zone'. However in
  the case of GX~339-4 ({\bf bottom panel}), a major radio flare event was observed to
  occur several days {\emph before} the r.m.s. drop, casting into doubt
  any direct causal connection between them. From \cite{fhb09}.}
\label{zones}
\end{figure}

\section{Connections to Active Galactic Nuclei}

\index{AGN}As the wealth of data on the coupling between accretion and ejection in
black hole X-ray binaries grew, more serious attempts were made to
scale physical properties up to the AGN. The binary studies had made
it clear that there was a strong dependence on accretion rate for a
given black hole (not surprisingly), and of course a strong dependence
of a variety of properties with black hole mass was also expected
(e.g. Fig. 2). Therefore some of the first attempts to quantitatively
scale properties between black hole binaries and AGN involved black
hole mass $M$, accretion rate $\dot{m}$ (or some proxy for it) and
some other property (namely radio luminosity or power spectral break
frequency). 

\subsection{Luminosity scalings}

\index{scaling}As noted above, Corbel et al. and Gallo et al. reported a non-linear
correlation between radio and X-ray luminosities in a number of hard
state black hole binaries. Shortly afterwards two groups (\cite{merloniheinz,falcke})
independently established the existence of a plane linking these
binaries with a large population of active galactic nuclei (AGN)
hosting supermassive black holes. The plane is based upon the relation
of the radio luminosity $L_{\rm radio}$, the X-ray luminosity $L_X$
and the black hole mass. This should be considered one of the major
steps in the unification of black hole accretion on all mass scales.

In the Merloni et al. formalism, the fundamental plane\index{fundamental plane} can be
represented as:

\[
L_{\rm radio} \propto L_X^{0.6} M^{0.8}
\]

where the power-law indices are fitted values to a large sample of
XRBs and AGN.

The most recent refinements of the plane are presented in \cite{gallo2006, koerdingfalcke}.
Criticisms of the
plane have been rebuked by a consortium of all the original discovery
authors, in \cite{merloni2006}.

Of the three parameters of the fundamental plane, one is genuinely
fundamental ($M$), and one is a good indication of the total radiative
output of the system and is therefore pretty fundamental ($L_X$).
However, the third parameter, $L_{\rm radio}$ is merely a tiny tracer
of the enormous power carried by the jets from these systems. The fact
that it seems to correlate so perfectly with $L_X$ is itself quite
amazing and indicates a remarkable stability and regularity in the jet
formation process. For example, for a X-ray binary in a bright hard
state, the radio emission can be estimated to constitute about
$10^{-7}$ of the total jet power, and yet over long timescales,
including phase of jet disruption and reformation, the correlation
between radio and X-ray luminosities holds very well (albeit not
perfectly -- S. Corbel, private communication). 

Returning to the plane, it might be a more useful indicator of
physical quantities and the flow of matter and power around the black
hole if $L_{\rm radio}$ could be replaced with, say, the total jet
power $L_J$ or the mass accretion rate $\dot{m}$. In fact we can do
both, based upon the relations established in \cite{koerdingfendermigliari} 
(hereafter KFM03; see also e.g. \cite{heinzgrimm,heinz2007}. 
In Fig. \ref{fp2} we use the relations
from this paper to `calibrate' the plane, by replacing the axes of
Merloni et al. with the physical quantities of jet power (left) and
accretion rate (right).

\begin{figure}
\includegraphics[width=12cm]{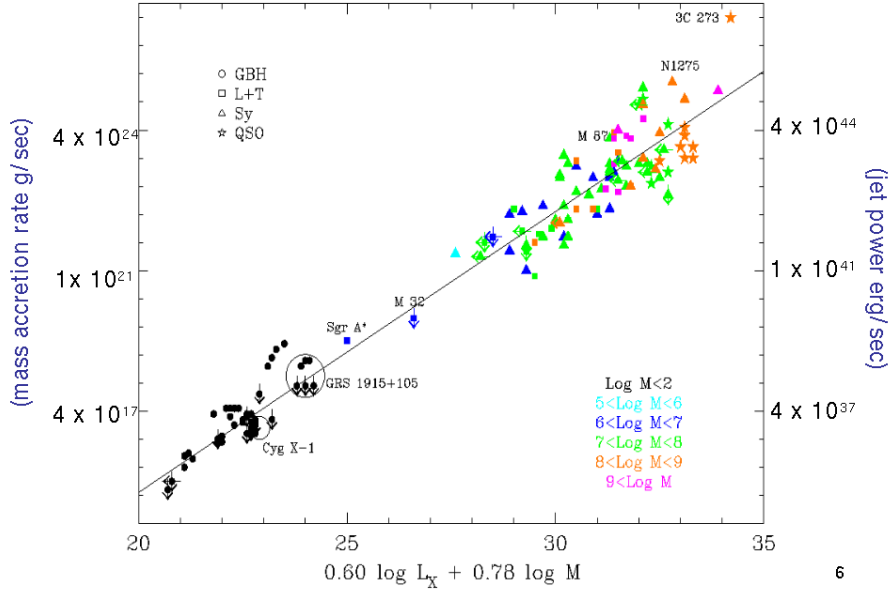}
\caption{The `fundamental plane'\index{fundamental plane} of black hole activity as presented
  originally in \cite{merloniheinz}, but in which their
  ordinate (y-axis) of radio luminosity has been replaced by the jet
  power and mass accretion rate (which scale linearly with each other)
  as estimated by \cite{koerding2007}.}
\label{fp2}
\end{figure}

\subsection{Variability scaling}\index{scaling}

\begin{figure}
\label{mch}
\includegraphics[width=12cm]{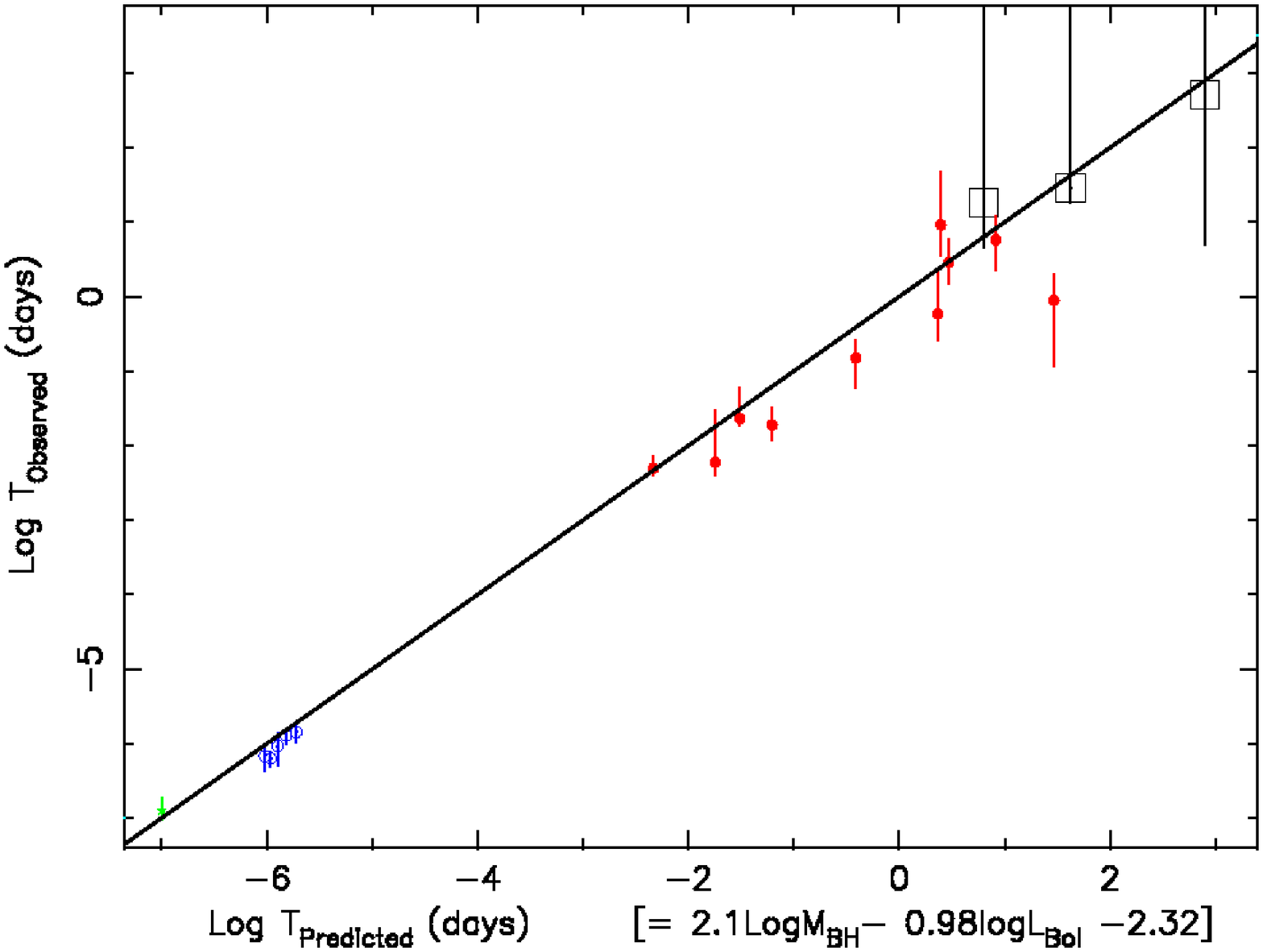}
\caption{A second fundamental plane, essentially relating
characteristic timescales to mass and accretion rate, from \cite{ian2006}. 
More recently \cite{koerding2007} have shown this to
extend `hard state' black hole X-ray binaries and neutron star
systems.}
\end{figure}

AGN X-ray timing studies started in the 1980s with the launch of
EXOSAT, which probed AGN X-ray variability on time-scales up to a few
days.  It was soon established that on these time-scales AGN showed
red-noise (i.e. ``1/f'') variability with steep unbroken PSDs
(\cite{lawrence,ian1987}).  However, McHardy (1988) noted using sparse
long-term archival data that the PSDs appeared to flatten or break on
longer time-scales, similar to the PSD shapes seen in BH XRBs.  The
launch of RXTE in 1995 allowed long-term light curves to be obtained
with extremely good sampling and categorically proved the existence of
PSD breaks on time-scales close to those expected by scaling the BH
XRB break time-scales up by the AGN BH mass
(e.g. \cite{uttley,markowitz,ian2004}), although with some
considerable scatter in the mass-time-scale relation.

Recently, \cite{migliari2005} established that for
a small sample of X-ray binaries there was a positive correlation
between radio luminosity and the frequencies of timing features. This
was to be expected, since we are confident that on the whole both
timing frequencies and radio luminosity are increasing functions of
accretion rate (although we also know there are other, state-related,
dependencies at high accretion rates).

\cite{ian2006} have now fitted a plane which relates mass
{\emph and} accretion rate to the break frequency in X-ray power spectra, such
that

\[
T_{\rm break} \propto M^{-2.1} L_{\rm bol}^{-1}
\]

Where $T_{\rm break}$ is the break timescale, reciprocal of the break
frequency, $M$ is black hole mass and $L_{\rm bol}$ is bolometric
luminosity (Fig 10). All of the sources used in the correlation are
believed to be in radiatively efficient states and so $L_{\rm bol}$ is
used as a proxy for accretion rate. Using this substitution converted
to accretion rate and integer power-law indices (i.e. $2 \sim 2.1$),
we arrive at

\[
T_{\rm break} \propto M / (\dot{m} / \dot{m}_{\rm Edd})
\]

revealing the expected linear correlation of break timescales with
black hole mass, albeit for a fixed Eddington ratio of accretion
rate. 

\cite{koerding2007} have extended and refined this variability
plane to include lower-luminosity black hole binaries and neutron star
binaries (and even considers white dwarf accretors in cataclysmic
variable binaries). Fig. \ref{k07} plots the extended plane; note that
the two lines indicate that a further parameter, related to X-ray
state, should be required to fit all sources (as we see in X-ray
binaries that timing frequencies can vary at the same X-ray
luminosity, if the spectral state is changing).

\begin{figure}
\includegraphics[width=12cm]{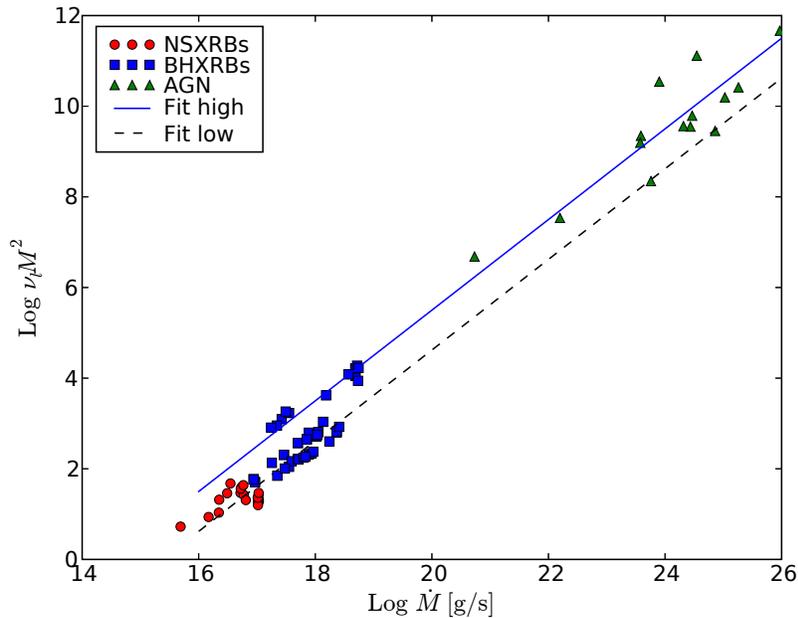}
\caption{The black hole timing plane extended to lower luminosity
  black holes {\emph and} neutron stars. From \cite{koerding2007}. The
  two lines hint at a further parameter, related to X-ray state, which
  may be required to fully fit all sources.}
\label{k07}
\end{figure}

\subsection{Further similarities}

As noted earlier, black hole X-ray binaries seem to (nearly) always
follow a pattern of behaviour in outburst similar to that sketched in
\cite{fbg,hombel} (see Fig.
6). However, it is clear that between different outbursts of the same
source, or outbursts of different sources, the luminosities at which
the hard $\rightarrow$ soft and soft $\rightarrow$ hard state
transitions may occur can vary quite significantly (e.g. Fig. 4; \cite{belloni,klein2008}).
As a result, an
ensemble of X-ray binaries would present a pattern in the
hardness:luminosity diagram with a long handle and a filled-in
head. Such an ensemble is obviously what we're going to have to deal
with if we want to be able to compare patterns of disc:jet coupling in
XRBs with those in AGN.

In \cite{koerdingjester2006} we have attempted to do
this. First we constructed the Disc Fraction - Luminosity Diagram,\index{Disc Fraction - Luminosity Diagram (DFLD)} in
which hardness is replaced by the ratio of power law to total
luminosity, a number which approaches zero for disc dominated soft
states, and unity for hard states. This is necessary for a physical
comparison, since the accretion discs temperature is a decreasing
function of black hole mass, and for AGN does not contribute
significantly in the X-ray band. We then simulated an ensemble of
BHXRBs, based upon \cite{fbg} and the slight
refinement (suggested in \cite{belloni}) that the `jet line'
might be diagonal in such a diagram. This was then compared to a
sample of AGN from the SDSS DR5 for which there were X-ray detections
and either radio detections or limits, {\emph plus} a sample of
low-luminosity AGN (LLAGN). The similarity was striking (see Fig.
\ref{turtles}) and suggests that {\emph the radio loudness is determined
by the combination of `state' and luminosity in a similar way for
accreting black holes of all masses}. Note that, while it is tempting
to consider, the diagram does {\emph not} indicate that AGN necessarily
follow the same anti-clockwise loop in the diagram as XRBs: the motion
in such loops could possibly be dominated by disc instabilities which
may not apply to AGN (although in the Fender, Belloni \& Gallo
interpretation, major radio flares would require right $\rightarrow$
left transitions in order to produce the internal shocks). What it
does indicate is that when an AGN finds itself in a particular
accretion `state', whether disc or corona dominated or some mix of the
two, the jet it produces will be comparable to that which a XRB would
make in the same state. \cite{kaiser} further discuss possible
similarities and differences between feeding cycles in AGN and X-ray
binaries.

\begin{figure}
\includegraphics[width=12cm]{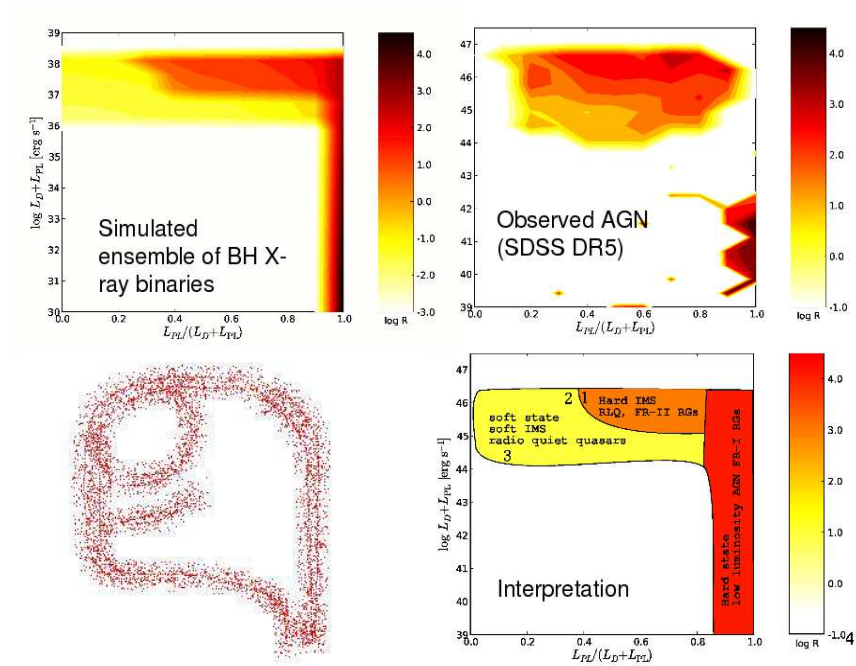}
\caption{A comparison of the disc:jet coupling in X-ray binaries with
that in AGN, based upon the Disc Fraction Luminosity Diagram (DFLD;
\index{Disc Fraction - Luminosity Diagram (DFLD)}
see \cite{koerdingjester2006}). Based upon the disc:jet model
for black hole X-ray binaries presented in \cite{fbg},
an example track for which is shown in the {\bf lower left panel}, a
simulated ensemble of X-ray binaries was produced ({\bf upper left
panel}). This was compared with a sample of SDSS quasars and low
luminosity AGN ({\bf upper right panel}) and a striking similarity
revealed. The lower right panel offers an explanation for the
similarities between the different classes of object.}
\label{turtles}
\end{figure}

It is worth noting that although the DFLD was chosen to both provide
easy and understandable comparison with the HID {\emph and} to provide a
clear method of physical comparison with accretion states in AGN, it
does show some differences with the HID even for X-ray binaries. 

Firstly, the work of \cite{dunn} who have produced the first
real DFLDs for an X-ray binary (GX~339-4) shows that the hysteretical
zone of the HID becomes rather less square in shape when transferred
to the DFLD. This seems to be because as the disc cools on the soft
branch, the disc fraction is dropping (not as obvious as it sounds --
it depends upon how fast the hard component is also
dropping). Secondly, the work of \cite{jeff1995,jeff2003} has
demonstrated that by the time sources are in quiescence, the radiative
luminosity of the disc is one to two orders of magnitude {\emph greater}
than the X-ray luminosity -- i.e. at the lowest accretion rates discs
are once more dominant in terms of the radiation. Of course at these
low accretion rates we still estimate that the kinetic power of the
jet dominates over both of these terms. Fig. \ref{dfld} presents the
DFLD for a set of black hole binaries for which the disc component
strength has been (relatively) well measured either in the soft X-ray,
ultraviolet or optical bands (from \cite{cabanac}). The
(designed) similarity with the HID at high luminosities is apparent,
but interestingly the path of a source loops back to the
disc-dominated state in quiescence. It is not clear that we would
expect to see this lower disc-dominated branch in AGN since it rather
reflects the conditions of the outer accretion flow. 

\begin{figure}
\includegraphics[width=12cm]{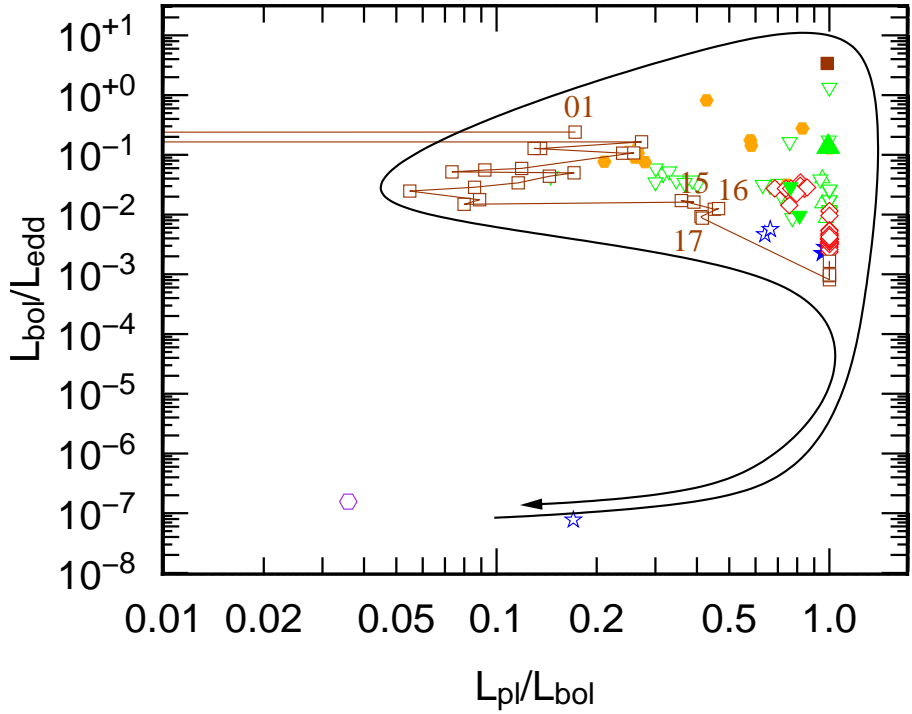}
\caption{The disc-fraction luminosity diagram (DFLD)\index{Disc Fraction - Luminosity Diagram (DFLD)} for black hole
  X-ray binary systems where the disc component is well measured in
  the X-ray band ({\emph Swift} or SAX) or optical bands. The connected
  brown points correspond to the {\emph Swift} observations of XTE
  J1817-330. The black line with arrowhead indicates a possible path
  of a transient black hole from quiescence to outburst and back
  again. It is well established, but not widely appreciated, that as
  well as in soft X-ray states, in quiescence systems are also
  completely dominated in their radiative output by the accretion disc
  (see \cite{jeff2003}). From \cite{cabanac}; all X-ray
  data have been independently reanalysed, while optical fluxes are
  from the literature.}
\label{dfld}
\end{figure}

Further similarities in patterns of behaviour have been noted in the
past. In Fig. \ref{marscher} patterns of X-ray temporal and spectral
behaviour are related to directly imaged relativistic ejection events
in the AGN 3C~120 (\cite{marscher2002}; see also Chap. 7).

\begin{figure}
\includegraphics[width=11cm]{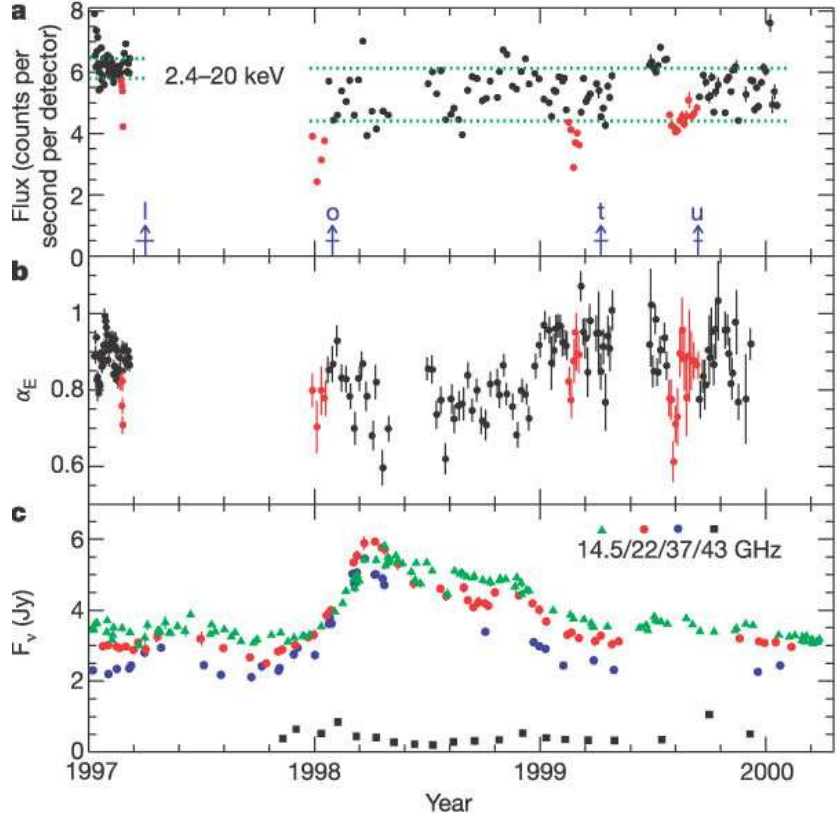}
\caption{Possible disc-jet coupling in the AGN 3C~120, from \cite{marscher2002}. 
{\bf Top panel:} X-ray light curve in photon counts per
second per detector; for many data points, the error bars are smaller
than the symbols. Times of ejections of superluminal radio knots, as
resolved with the VLBA, are indicated by upward blue arrows, with the
horizontal crossbar giving the uncertainty. {\bf Middle panel} X-ray
hardness; lower value is harder.{\bf Lower panel} Radio light curves
at 14.5 GHz (green triangles); 22 GHz (red circles) and 37 GHz (blue
circles); and 43 GHz (core only, black squares).}
\label{marscher}
\end{figure}

\subsection{Differences}

Despite the clear similarities, there are clear differences, both
expected and observed, between X-ray binaries and AGN. Some of these
result from the fact that these systems are more than just the central
black hole. In particular the {\emph environment} may play a role in 

\begin{enumerate}
\item{supplying matter with varying distributions of angular momentum,
  possibly even changing direction of rotation, depending on the
  merger history of the host galaxy and the path taken by the gas in
  reaching the region of the black hole (e.g. \cite{emse} and
  references therein). This clearly does not occur in Roche-lobe
  overflow binary systems, in which mass and angular momentum transfer
  takes place at a more or less constant rate.}
\item{Obscuration and inclination play major roles in the appearance
  of AGN (e.g. \cite{urry}; however these authors do not
  really consider the relation of appearance to accretion rate /
  state). There may be some parallels with this in the case of some
  X-ray binaries, but in the majority it is not the case.}
\item{Jets from AGN in dense environments will probably dissipate
  their energy on smaller physical scales as they inflate the medium
  around them. There may be some parallel here between e.g. GHz-peaked
  AGN (e.g. \cite{odea}) and the interaction of powerful jets and
  strong stellar winds in the systems like Cygnus~X-3 and SS~433.}
\end{enumerate}

In addition to environmental effects, there is (at least) one key
region in the geometry of AGN which appears simply not to be present
in X-ray binary systems: the Broad Line Region (BLR)\index{Broad-Line Region}. The broad
emission lines in AGN have been shown by reverberation mapping
techniques to arise within typically a few days of the central black
hole (\cite{peterson2004} and references therein). The precise
velocity field of the BLR is however uncertain. \cite{elvis} presents
a clear case for a BLR that originates in a relatively thin conical
outflow which has its origin in the accretion disc, but whose major
component of motion is away from the disc.  However,
\cite{peterson1999} argue that the variable components to BLR emission
lines show the expected virial relation between velocity $V$ and
radius $r$, $V \propto r^{-1/2}$, implying that the motions of the
line-emitting regions are dominated by the gravity of the central
black hole. This facilitates the use of reverberation mapping
techniques to estimate central black hole masses (\cite{peterson2004}
and references therein, but see \cite{krolik2001} for a critique of
the method).  Other works, e.g. \cite{wanders}, also conclude that the
motion in the BLR is primarily
circular/orbital. \cite{murray1996,murray1997,murray1998} have studied
in detail the appearance of emission lines from combined accretion
disc plus wind systems, with application to both binary systems
(cataclysmic variables) and AGN, and have shown that single-peaked
lines can arise from rotating flows.  Why then is there no clear
evidence for an outflowing BLR in black hole X-ray binaries, when
there are so many other apparently scale-free similarities in black
hole accretion and jet formation ?

Firstly, \cite{proga} have already demonstrated, on the
basis of detailed calculations, a strong line-driven wind is unlikely
in the case of accretion discs around X-ray binaries. The problem is
essentially that the central X-ray source is too strong a source of
ionising radiation, which strongly inhibits line-driving. In AGN, in
contrast, the cooler central disc is a UV source which is suitable for
driving such winds. What if there are significant winds driven off by
other means such as large-scale magnetic fields (e.g. \cite{blandford,miller2006} 
or thermal expansion \cite{begelman}? In this case we can ignore the line-driving
requirement. Nevertheless, the strongly ionising X-ray source still
prevents line emission from the inner, fast moving, regions of this
wind. For example, the BLR for a $10^{8}$ M$_{\odot}$ black hole in an
AGN, accreting at close to the Eddington limit is typically estimated
to be at a distance of several tens of light days ($10^{16}$ cm $\leq
R_{\rm BLR} \leq 10^{17}$ cm), or 300--3000 gravitational radii
\cite{peterson2004}. The line-emitting region of the disc is fairly
close to this \cite{dultzin}. On the other hand,
X-ray to optical delays in BHXBs accreting at comparable Eddington
ratios (e.g. \cite{hynes2006}) indicate lags of order seconds, or $>
10^4$ gravitational radii. The difference in the radii of the
line-emitting regions of the accretion discs is due, once again, to
the much hotter environments around stellar-mass black holes in BHXBs
($T \propto M^{-1/4}$ at the same Eddington ratio, and for a
radiatively efficient flow). Therefore it seems likely, albeit in this
case without detailed calculations, that even if a strong wind could
be launched without line-driving, if it were launched from close to
the black hole then the gas would be too highly ionised to produce
BLR-like lines. 

\subsection{What about spin ?}\index{black hole spin}

There has been much speculation over the past few decades about the
possible role of black hole spin in jet formation (e.g. \cite{wilson,koide}) 
and any relation it might have to the
apparent radio loud:radio quiet dichotomy in AGN (listed as one of the
ten major questions for the field by \cite{urry}). There
have, more recently, been strong claims for clear measurements of
black hole spin in AGN (e.g. \cite{wilms}) and X-ray binaries
(e.g. \cite{miller2002}). It is this authors view that to date there
is no clear evidence either way for the role of spin-powering of jets
in black hole X-ray binaries (contrary to the view put forward in
e.g. \cite{zhang}). Some points to consider, however, are

\begin{itemize}
\item{In black hole binaries both `radio loud' and `radio quiet'
states are observed from the same source, repeatedly, and are related
to the accretion state, not spin changes}
\item{In AGN there may be a larger range of black hole spins, relating
to the merger history, whereas all black holes in binary systems have
(presumably) formed from the collapse of a massive star and will have
some intermediate value of spin}
\end{itemize}

\section{Using, testing and exploring}

Now that we have established the first good quantitative scalings, as
well as a new set of qualitative similarities, between black holes on
all mass scales, what do we use them for ? The goal expressed by many
for several years was to use the insights we gain from the disc--jet
coupling in stellar-mass black holes to understand better the process
of accretion and feedback in AGN. The importance of this connection
has been greatly strengthened in recent years by the discovery that
feedback from black holes is likely to be directly linked to both the
formation of galaxy bulges \cite{gebhardt,ferrarese}, 
and to the heating of the inner regions of cooling flows
(e.g. \cite{best,vernaleo}).

An obvious and important place to start is therefore with the kinetic
feedback of AGN, or kinetic luminosity function. \cite{merloni2007} 
and, more directly, \cite{koerdingjester2007} have
estimated the kinetic luminosity function of AGN based upon both power
and spectral state scalings from black hole binaries. Both groups
conclude that kinetic feedback into the IGM is actually likely to be
dominated by supermassive black holes accreting at relatively low
Eddington ratios. Future studies need a better regulated sample with
mass measurements and more precise measurements of core coronal, disc
and jet components. 

Another example relates to the argument originally noted by \cite{soltan}
(see also e.g. \cite{fabian} and references therein), that the
local mass density of black holes is consistent with the growth of
black holes via {\em{radiatively efficient}} accretion (current limits
place a factor of a few on this consistency). However, as is clearly
the case for binary black holes, radiatively efficient accretion above
$\sim 1$\% Eddington, can occur with powerful jets (hard and hard
intermediate states) or without a jet (soft state). Taking extremes,
if most of the X-ray background results from accretion in hard or hard
intermediate states, then $\sim 10^{67}$ erg Gpc$^{-3}$ may have been
injected into the ambient medium in the form of kinetic energy over
the lifetime of black hole growth. If most of the X-ray background
results from accretion in jet-free {\emph soft} X-ray states then the
figure will be much smaller, probably by a factor of a hundred or
more.

\subsection{Somewhere in the middle: using radio emission to look for 
intermediate mass black holes}\index{intermediate mass black holes}

The publication of the fundamental planes of black hole activity by
\cite{merloniheinz} and \cite{falcke} coincided with a period
of greatly renewed interest in `ultraluminous' X-ray sources (ULXs)
\index{Ultraluminous X-ray Sources (ULX)}
and the suggestion that some of these object could be `intermediate
mass' black holes ($10^2 M_{\odot} \leq M \leq 10^4 M_{\odot}$). These
ULXs were not located at the dynamical centres of their host galaxies,
and could sometimes produce X-ray luminosities in excess of $10^{41}$
erg s$^{-1}$, or the Eddington limit for a $\geq 70 M_{\odot}$ black
hole, {\emph if the X-ray emission was isotropic}. Alternatives based
upon anisotropy of the emission, whether intrinsic or associated with
relativistic aberration, have been put forward (e.g. \cite{king};
\cite{koerding2002}.).

The fundamental planes immediately implied two things which were
potentially good for this field:

\begin{enumerate}
\item{If you could measure the X-ray and radio luminosities of a ULX, you could infer its mass}
\item{Radio observations were the best way to find intermediate mass black holes accreting at low rates from the ambient medium}
\end{enumerate}

Point [1.] has had mixed success. Several ULXs do appear to have radio
counterparts, and simply plugging these radio luminosities into the
fundamental plane implies large black hole masses, $>> 100
M_{\odot}$. However, in most cases this radio emission is in fact
resolved and looks like a large scale nebula (e.g. \cite{mushotsky,soria}),
and no very compact and variable radio
counterparts have yet been found.

Point [2.] has been explored in some detail by \cite{maccarone2005} and
\cite{maccarone2005al} (see also\cite{maccarone2008}).
Put simply, black holes accreting at low rates should be easier
to find in the radio band. This situation is enhanced the greater the
mass of the black hole, as the radiative efficiency seems to fall off
with Eddington ratioed accretion rate. Recent confirmation of this
approach may have come in the form of the detection of a radio source
at the centre of the globular cluster G1 by \cite{ulvestad},
which is consistent with a black hole of mass $\geq 100
M_{\odot}$.

\subsection{Comparison with neutron stars and white dwarfs}

\index{neutron stars}\index{white dwarfs}A control sample exists which allows us to test whether properties
unique to black holes, such as the presence of an event horizon, are
in any way essential for any of the observed phenomena. Neutron stars
are collapsed stellar remnants with masses $1 M_{\odot} \leq M_{\rm
NS} \leq 2 M_{\odot}$ and sizes only a factor of 2--3 larger than
their Schwarzschild radii (see Fig. \ref{sizes}). Squeeze them to make
them half as large as they are and they would collapse to form black
holes. The gravitational potential energy released per unit mass of
matter accreted onto a neutron star is therefore very similar to that
of a black hole. In addition, we find them in X-ray binary systems
with very similar patterns of accretion, including outbursts and
extended periods of quiescence, to the black hole binaries. A full
review of the properties of jets from neutron star binaries is beyond
the scope of this paper, but the key points from the comparison are

\begin{enumerate}
\item{Neutron stars seem to produce both steady and transient jets,
just like black holes, with a similar relation to hard states and
outbursts (although poorly sampled to date)}
\item{The radio to X-ray ratio is in general lower for neutron star binaries}
\item{Neutron star jets are less `quenched' in soft states than black hole jets}
\item{Neutron star jets may be just as, or even more, relativistic (in
terms of bulk velocity) than black hole binary jets}
\item{The radio:X-ray correlation, although much less well measured,
appears to be steeper than that for black holes by a factor of two or
more, i.e. $L_{\rm radio, NS} \propto L_{\rm X, NS}^{\geq 1.4}$
compared to $L_{\rm radio, BH} \propto L_{\rm X, BH}^{\sim 0.7}$}
\item{Two neutron star jet systems (Sco X-1, Cir X-1; possibly also
the `odd' system SS 433) appear to show unseen but highly relativistic
flows which energise slower-moving bulk flows further out.}
\end{enumerate}

\cite{migliarifender} present the most comprehensive review to
date of the properties of jets from neutron star binaries (see also
\cite{koerdingfendermigliari} and Chap. 4 in this
volume, and references therein).

Several of the points listed above strongly imply that the basics of
relativistic jet formation do not require any unique property of black
holes, such as an event horizon. However, as discussed earlier (and
illustrated in Fig. \ref{sizes}), in terms of gravitational potential,
neutron stars are almost black holes, and so perhaps the similarities
are not so surprising. However, there is another major class of
accreting binary systems, which contain accreting objects which are
very different: {\emph Cataclysmic Variables} (CVs). In these systems a
white dwarf, with a $M / R$ ratio about a thousand times smaller than
for a black hole or neutron star, accretes matter from a companion
(white dwarf accretion is less efficient than nuclear fusion).

However, \cite{koerding2008} have recently shown that there may
even be direct similarities between accretion state changes at high
luminosity in disc-accreting CVs and black holes and neutron
stars. Fig. \ref{sscyg} presents HIDs or equivalent for the black hole
binary GX~339-4\index{ GX~339-4}, the neutron star binary Aql~X-1\index{Aql~X-1} and the disc-accreting
CV SS~Cyg\index{SS~Cyg}. The similarity is striking, but what is key is that
observations with the VLA have revealed radio flaring at around the
point of the high luminosity state transition in SS~Cyg, just as seen
in GX~339-4 and other black hole binaries. This strongly suggests the
transient formation of a jet during this phase, further suggesting
that {\emph patterns} of disc:jet coupling relate solely to the
behaviour of the disc and not the central accretor (which may,
however, affect e.g. jet velocity).

\begin{figure}
\includegraphics[width=10cm, angle=270]{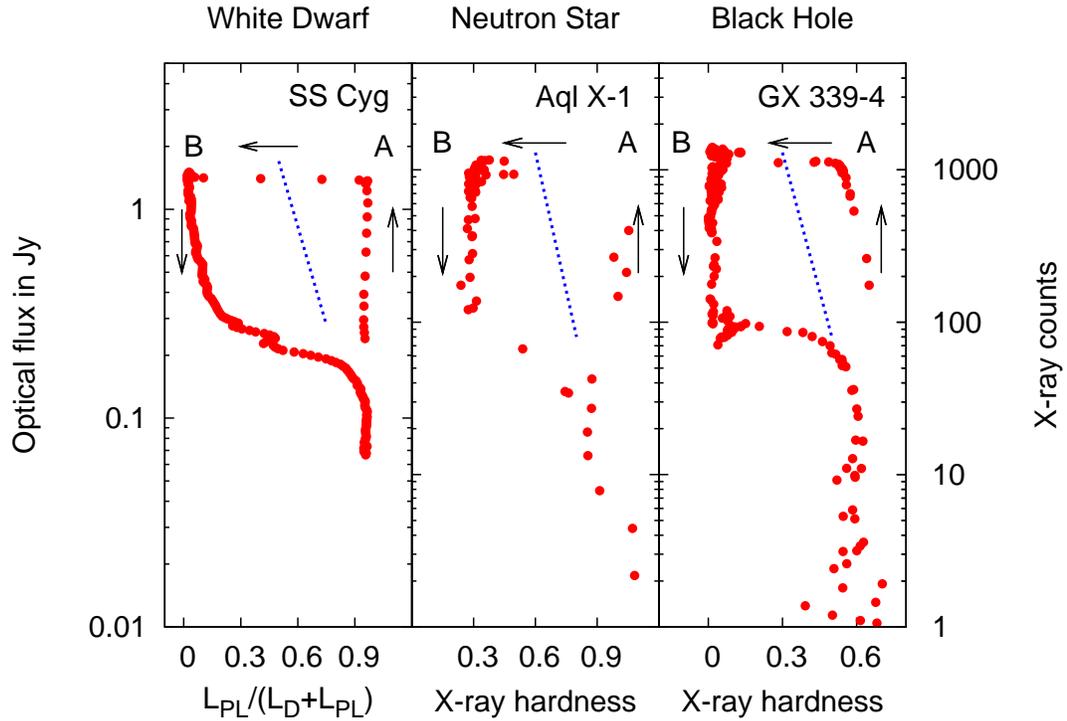}
\caption{HID for a black hole, a neutron star, and the disc-accreting
CV SS~Cyg. The arrows indicate the temporal evolution of an
outburst. The dotted lines indicate the jet line observed in black
hole and neutron star XRBs: On its right side, one generally observes
a compact jet; the crossing of this line usually coincides with a
radio flare. For SS~Cyg, we show a disc-fraction luminosity
diagram. We plotted optical flux against the power-law fraction
measuring the prominence of the "power-law component" in the hard
X-ray emission in relation to the boundary layer/accretion disk
luminosity. This power-law fraction has similar properties to the
X-ray hardness used for XRBs. Radio flaring has been observed during
the high luminosity state transition in SS~Cyg just like in black hole
systems.}
\label{sscyg}
\end{figure}

To conclude, there are clear similarities, both quantitative and
qualitative, between accretion (luminosity, state) and jets
(formation, steady or transient) in black holes of all masses. These
links provide us with tools to understand feedback from AGN and its
role in the formation of galaxies and evolution of clusters, and with
insight into the process of relativistic jet formation. However, we
should be cautious about drawing wide-ranging conclusions about the
effect of black hole properties, e.g. spin, on jet formation, when we
find comparable patterns of behaviour in related (neutron star) and
also very different (white dwarf) classes of object.

\subsection{Further reading}

This review has necessarily been both brief and biased, and the
literature associated with this area of research is vast and
ever-expanding. In particular, I have approached the area of disc--jet
coupling from the realm of X-ray binaries, looking for scalings and
comparisons with AGN, and have not even scratched the surface of the
wide phenomenology associated with accreting supermassive black holes.

For a broader introduction to AGN, I would recommend \cite{krolik1999} and
references therein. For a more AGN-centric view of the disc-jet
coupling around accreting black holes, see papers by Marscher,
e.g. \cite{marscher2006}. For an up-to-date reference list of all things
related to black holes, see \cite{gallo2008}.

\subsection*{Acknowledgements}

I owe much of my understanding of this field of research to
collaborations and conversations with a large number of people, most
notably those who are co-authors with me on a wide range of
papers. They know who they are, and if {\em{you}} don't, use ADS. In
addition I would like to thank the participants at the ISSI workshop
in January 2008 for helping to further clarify ideas, and Christian
Knigge, Mike Goad and Daniel Proga for discussions about broad line
regions and line-driven winds.

\end{document}